\pdfoutput=1
\documentclass{article}
\usepackage{arxiv}
\usepackage{babel}
\usepackage{xcolor}
\usepackage{graphicx}
\usepackage{amsmath,amsfonts}
\usepackage{graphicx,indentfirst,epsfig}
\usepackage{varioref}
\usepackage{wrapfig}
\usepackage{subcaption}
\usepackage{amsthm}
\usepackage{bm}

  \def\clap#1{\hbox to 0pt{\hss#1\hss}}

\providecommand{\mat}[1]{\bm{#1}}%
\renewcommand{\vec}[1]{\mathbf{#1}}

\providecommand{\mA}{\ensuremath{\mat{A}}}
\providecommand{\mB}{\ensuremath{\mat{B}}}

\providecommand{\mE}{\ensuremath{\mat{E}}}

\providecommand{\mI}{\ensuremath{\mat{I}}}

\providecommand{\mK}{\ensuremath{\mat{K}}}

\providecommand{\mV}{\ensuremath{\mat{V}}}
\providecommand{\mW}{\ensuremath{\mat{W}}}
\providecommand{\mX}{\ensuremath{\mat{X}}}

\providecommand{\mSigma}{\ensuremath{\mat{\Sigma}}}

\providecommand{\va}{\ensuremath{\vec{a}}}

\providecommand{\vf}{\ensuremath{\vec{f}}}

\providecommand{\vv}{\ensuremath{\vec{v}}}

\providecommand{\vx}{\ensuremath{\vec{x}}}
\providecommand{\vy}{\ensuremath{\vec{y}}}

\usepackage{hyperref}
\usepackage{cleveref}
\usepackage{scrextend}
\usepackage{algorithm}
\usepackage{algorithmic}
\usepackage{booktabs}
\usepackage{longtable}
\usepackage{soul}
\usepackage{lineno}

\title{Prior-informed uncertainty modelling with Bayesian polynomial approximations}
\date{}
\author{ Chun Yui Wong\thanks{Corresponding author} \\
	Department of Engineering\\
	University of Cambridge\\
	Cambridge, United Kingdom \\
	\texttt{ncywong@cantab.net} \\
	\And
	Pranay Seshadri \\
	Department of Mathematics (Statistics Section)\\
	Imperial College London\\
	London, United Kingdom \\
	\And
	Andrew B. Duncan \\
	Department of Mathematics (Statistics Section)\\
	Imperial College London\\
	London, United Kingdom \\
	\And
	Ashley Scillitoe \\
	Data Centric Engineering\\
	The Alan Turing Institute\\
	London, United Kingdom \\
	\And
	Geoffrey T. Parks \\
	Department of Engineering\\
	University of Cambridge\\
	Cambridge, United Kingdom \\
}

\begin{document}

\maketitle

%
%
%
%
%
%
%
%
%

\begin{abstract}
Orthogonal polynomial approximations form the foundation to a set of well-established methods for uncertainty quantification known as polynomial chaos. These approximations deliver models for emulating physical systems in a variety of computational engineering applications. In this paper, we describe a Bayesian formulation of polynomial approximations capable of incorporating uncertainties in input data. Through different priors in a hierarchical structure, this enables us to incorporate expert knowledge on the inference task via different approaches. These include beliefs of sparsity in the model, approximate knowledge of the polynomial coefficients (e.g., through low-fidelity estimates) or output mean, and correlated models that share similar functional and/or physical behaviours. We show that, through a Bayesian framework, such prior knowledge can be leveraged to produce orthogonal polynomial approximations with enhanced predictive accuracy.
\end{abstract}

\keywords{Polynomial chaos, Bayesian inference, uncertainty quantification, coregional models}

\section{Introduction}
Recent years have heralded the development and deployment of uncertainty quantification methods throughout computational physics. Arguably, polynomial chaos (PC) \cite{xiu2002wiener} has been one of the gems that has come out of this global push towards greater rigour in working with simulations. PC is the utilisation of a weighted sum of orthogonal polynomials to construct a representation of a computational model. This model may take the form of a black box---where only inputs and outputs are known---or a set of partial differential equations---where analytic forms of the underpinning equations are available---constituting the non-intrusive \cite{xiu2007efficient} and intrusive \cite{pettersson2015polynomial} branches of PC respectively. In both branches, certain input parameters are assumed uncertain and are duly assigned probability distributions. The impact of this uncertainty on the physical system emulated can be propagated to assess what the mean, variance, and sensitivities in the outputs are. PC affords straightforward estimates of these quantities, and as a result has seen tremendous practical uptake across the engineering \cite{prabhakar2010polynomial, witteveen2007modeling, ghanem1990polynomial, spiridonakos2016polynomial} and science \cite{wang2021development,kalra2017sensitivity, jones2013nonlinear} communities. 

In parallel to its industrial use, research into PC (and more broadly orthogonal polynomials) has been burgeoning. Key topics of interest include (i) data-driven dimension reduction \cite{hokanson2018data, constantine2017near}, (ii) deterministic and non-deterministic sampling approaches and numerical quadrature \cite{narayan2017computation, seshadri2017effectively}, (iii) methods for tackling correlations in the input parameters \cite{jakeman2019polynomial}, and (iv) the computing of higher-order moments and sensitivities \cite{wong2019extremum, geraci2016high, sudret2008global}. 

It is worth noting why PC has been so successful. First, it is purpose-built for negotiating uncertainties in physics-based models where the inputs are typically boundary conditions and geometry parameters, while the output quantities of interest (QoIs) are often spatio-temporal integrals of scalar field quantities, obtained from the governing partial differential equations. The functional behaviour between these inputs and outputs is usually smooth and continuous, even if the underlying scalar fields themselves exhibit discontinuities \cite{wong2020embedded}, e.g., shocks in Euler flow. Second, we tend to trust our models within a relatively small space defined by their geometry parameters and boundary conditions, where output QoIs are not extremely non-linear. This makes polynomials ideal candidates for characterising such input-output maps. Third, even for problems with several dimensions, PC can offer favourable estimates of output QoI moments with few model evaluations (in the case of non-intrusive PC). Research into more efficient sampling approaches that can deliver well-conditioned linear systems has facilitated a shift away from expensive tensorial and even sparse grid sampling techniques. To industry, this amounts to more rapid design cycles with limited high performance computing expenditures. 

The paragraphs above offer a brief overview of the state of PC today. While it is the answer to many aleatory uncertainty quantification challenges, it is not capable of addressing some of the emerging trends seen in computational physics. Below we identify some of these salient issues.

\begin{itemize}
\item One of the most pressing is the ingestion of the epistemic uncertainty associated with a single evaluation of a numerical simulation at a particular set of input boundary conditions. This uncertainty, which arises from assumptions in the governing physical model---i.e., length scales considered, numerical discretisation adopted, turbulent phenomenon modelling, temporal averaging, partial differential equations used and assumptions therein, is prevalent across both fluids \cite{duraisamy2019turbulence} and structural \cite{girolami2019statistical} domains. At present there is no mechanism within PC to account for an uncertainty in each model evaluation. Existing approaches to address this have been via the use of the worst-case solution from an epistemic uncertainty analysis as the model evaluation in the standard aleatory PC workflow \cite{emory2016uncertainty}, encapsulation approaches \cite{jakeman2010numerical}, or through interval analysis \cite{swiler2009epistemic, terejanu2010approximate}.
\item In PC, the expert modeller typically has no interactive role to play once the simulation pipeline has been set up. Thus, even when they have a \emph{good feel} for what the model output is, or even where the model can be trusted more, that insight is not passed to the PC workflow. Even in the absence of a rigorous framework to quantify epistemic uncertainty, a na\"{i}ve strategy for conveying a modeler's certainty in an output QoI instance should be captured. Such a human-in-the-loop approach has had significant impact in the broader field of machine learning \cite{budd2021survey}, but limited work has been done in this area of uncertainty quantification.
\item There is undoubtedly far greater awareness today of the misalignment between physical hardware testing and computational engineering. Most numerical simulations are validated on canonical cases (e.g., airfoils, beams, channels), but deployed on cases with far greater complexity (e.g., aircraft, bridges, jet engines). For a variety of reasons---that fall under the colloquial expressions known knowns, known unknowns, and unknown unknowns \cite{rumsfeld}---numerical simulations often fail to match their experimental counterparts, particularly for complex geometries, flows, topologies, and previously unseen physics \cite{denton2010some}. Developing a mathematical framework where PC models can leverage both experimental data and computational simulations concurrently would provide a roadmap for digital twinning efforts.
\end{itemize}

In an attempt to address these points, we consider a Bayesian approach to PC, an approach which has been the focus of some recent research efforts. For example, Ranftl and von der Linden \cite{ranftl2021bayesian} present a probabilistic method to quantify the uncertainty arising from the approximation error of surrogate models, and use a PC model as an example. Cheng, Lu and Zhen \cite{cheng2019multi} considered a Bayesian approach for multi-fidelity sparse PC approximations where the posterior polynomial is viewed as a Gaussian process (GP). We view our contributions within a much broader and unifying framework, offering a fully Bayesian treatment of PC, providing novel and useful tools for addressing epistemic uncertainty quantification, human-in-the-loop uncertainty quantification, and experimental vs.~numerical misalignment issues, among others. 

The rest of this paper is structured as follows: in \Cref{sec:bpc}, Bayesian PC (BPC) is introduced by re-interpreting classical ideas from PC by treating the solution of coefficients as a probabilistic problem. The design of a prior for the polynomial coefficients to transfer information both on its values (a physically-informed prior) and its sparsity pattern (a structural prior) is discussed in this section. This framework allows us to define the approximant as a stochastic process, e.g., in the case of Gaussian priors, a GP. The application of ideas from GP regression leads to methods where one can leverage prior information on output moments (\Cref{sec:moments}) and other similar models (\Cref{sec:coregional}) to improve prediction. In \Cref{sec:examples}, several analytical and simulation-based test cases are presented to illustrate the utility of BPC.

\section{Bayesian polynomial chaos} \label{sec:bpc}

\subsection{Preliminaries: polynomial chaos} \label{sub:prelim}
In classical PC, we consider the problem of approximating a QoI which is a function $f(\vx)$ of certain \emph{input parameters} $\vx = [x_1, x_2, ..., x_d]$. The input parameters are considered as a random vector taking values within a subset $\mathcal{D}$ of $\mathbb{R}^d$, which we assume can be written as the Cartesian product of one-dimensional (possibly infinite) intervals,
\begin{equation}
\mathcal{D} = \mathcal{D}_1 \times \mathcal{D}_2 \times ... \times \mathcal{D}_d,
\end{equation}
corresponding to each of the $d$ input parameters. These inputs are assumed to be independently distributed, and endowed with an \emph{input probability density function} $\boldsymbol{\rho}(\vx)$ which is the product of marginal distributions, 
\begin{equation}
\boldsymbol{\rho}(\vx) = \rho_1(x_1) \times \rho_2(x_2) \times ... \times \rho_d(x_d).
\end{equation}
The independence assumption can be relaxed, but we do not pursue this topic for simplicity of exposition. The input probability density function characterises the input uncertainty to the model $f(\vx)$. Under mild assumptions, classical results from PC show the existence of an approximation in the form of a linear combination of polynomial basis functions $\phi_j(\vx)$,
\begin{equation}
f(\vx) \approx g(\vx) = \sum_{j=1}^N \alpha_i \phi_j(\vx),
\end{equation}
where $\alpha_j$ are unknown \emph{coefficients}, for $j=1,...,N$. The size of the basis $N$ is termed the cardinality. Under the assumption of input independence, the basis polynomials are themselves products of univariate basis polynomials obeying an orthogonal relation, 
\begin{equation}
\phi_j(\vx) = \prod_{i=1}^d \phi_j^{(i)} (x_i), \qquad \int_{\mathcal{D}_i} \phi_j^{(i)} (x_i)~ \phi_k^{(i)} (x_i) ~\rho_i (x_i) ~dx_i = \begin{cases}
1 \quad \text{if } j=k,\\
0 \quad \text{otherwise.}
\end{cases}
\end{equation}
Consequently, the polynomial basis functions are identified by a multi-index $j = (j^{(1)}, j^{(2)}, ..., j^{(d)}) \in \mathcal{B} \subset \mathbb{N}^d$ denoting the degree of polynomial on each input axis. The set of possible multi-indices $\mathcal{B}$ is called the index set, with cardinality $N$. Common index sets include the tensor grid, total order, Euclidean degree, sparse grids and hyperbolic cross index sets, with the choice dictating the number of unknown coefficients, influencing the balance between expressiveness, ease of computation and generalisation capabilities. The choice of orthogonal polynomials for the task of approximating stochastic QoIs ensures exponential convergence when estimating moments of $f(\vx)$. These moments are also straightforwardly obtained from the coefficients, a point we emphasise in \Cref{sec:moments}.

Working under the framework of PC, the task of inference about the output QoI is underpinned by the estimation of the coefficients $\boldsymbol{\alpha} = [\alpha_1, \alpha_2, ..., \alpha_N]$ given a finite number of model evaluations captured by a set of $M$ \emph{input-output pairs}, $\{\mX, \vy\} = \{\vx^{(i)}, y^{(i)}\}$ where $y^{(i)} = f(\vx^{(i)})$ for $i=1,..,M$. Fixing the choice of basis functions, the task of coefficient estimation can be formulated as solving a parameterised matrix equation,
\begin{equation} \label{equ:PC_coeffs}
\widetilde{\vy} = \mV(\mX) \boldsymbol{\alpha},
\end{equation}
with $\mV(\mX)$ being a  \emph{weighted Vandermonde-type} matrix of the inputs, i.e., $\mV_{ij} = \omega_i \phi_j(\vx^{(i)})$ for some weights $\omega_i$, and $\widetilde{\vy}$ being the correspondingly weighted version of the output data $\vy$. The equality in \eqref{equ:PC_coeffs} takes on different interpretations. For example, assuming the number of data points $M$ is larger than $N$, a least-squares approach can be taken to solve \eqref{equ:PC_coeffs}. When $M<N$, strategies such as compressed sensing and ridge approximations can be used under certain assumptions about the output function.

\subsection{Bayesian polynomial chaos as a stochastic process} \label{sub:bpc_gp}
In this section and the subsequent ones, we depart from the standard PC workflow where \emph{exact} evaluations of the model are available. Instead, our observations $y$ are no longer equal to $f(\vx)$ for the intended input $\vx$, but are corrupted versions. Using the notation $y(\vx)$ for an observation corresponding to the input $\vx$, we can write
\begin{equation}
y(\vx) = f(\vx) + \varepsilon,
\end{equation}
where the error metric $\varepsilon$ is an unknown deviation from the truth. This quantity can encapsulate the epistemic uncertainty about the QoI inevitably introduced in the modelling process; for instance, the neglect of certain input factors and simplifying assumptions placed on numerical simulations of intractable physical phenomena such as turbulence. In this work, we make the simplifying assumptions that $\varepsilon$ can be modelled by a stationary (independent of $\vx$) and zero-mean Gaussian random variable,
\begin{equation} \label{eqn:noise_var}
\varepsilon \sim \mathcal{N} (0, \sigma_m^2),
\end{equation}
where $\sigma_m^2 > 0$ is the \emph{data variance}. With this assumption, we reformulate the task of coefficient estimation in a probabilistic framework. Treating the coefficients $\boldsymbol{\alpha}$ as random and using Bayes' rule, we can write 
\begin{equation}
 \textrm{p} \left( \bm{\alpha} |  \vy , \mV, \sigma_m^2 \right) \propto  \textrm{p}\left( \vy | \bm{\alpha} ,  \mV,  \sigma_m^2 \right) \times \textrm{p} \left( \bm{\alpha}\right),
\end{equation}
where the terms in the equation refer to the posterior, likelihood and prior of the coefficients, respectively. The likelihood is assumed to take the form of
\begin{equation} \label{eqn:likelihood}
\textrm{p}\left( \vy | \bm{\alpha} ,  \mV,  \sigma_m^2 \right) = \mathcal{N} (\mV \boldsymbol{\alpha}, \sigma_m^2 \mI),
\end{equation}
where $\mI$ is the identity matrix. In writing this down (instead of $\textrm{p}\left( \vy | \vf,  \sigma_m^2 \right)$), we make the assumption that the (deterministic) mismatch between $f(\vx)$ and $g(\vx) = \mV \boldsymbol{\alpha}$ is negligible. This source of error---also known as the \emph{truncation error} in classical PC---is intrinsic to the PC framework.  This is due to the finite dimensionality of the function space of polynomials, within which the polynomial approximant $g$ resides. The decision to neglect this mismatch relies on assumptions about the smoothness of the function, appropriate to most physical scenarios with some notable exceptions. 

\begin{figure}
\centering
\includegraphics[width=0.3\textwidth]{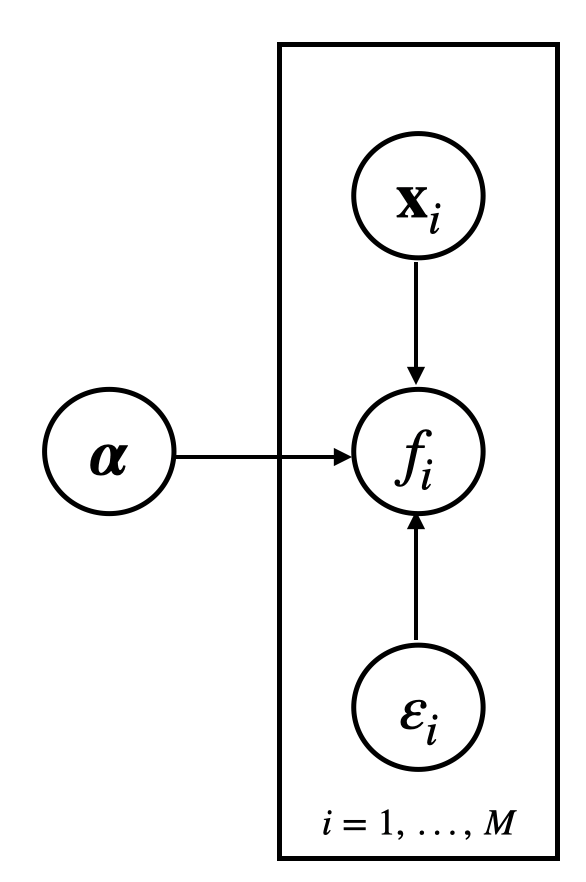}
\caption{Graphical model of Bayesian PC for $M$ input-output pairs.}
\label{fig:graph1}
\end{figure}

Further, making the assumption that the prior distribution $\textrm{p}(\boldsymbol{\alpha})$ is Gaussian, with
\begin{equation}
\textrm{p}(\boldsymbol{\alpha}) \sim \mathcal{N} (\boldsymbol{\mu}_\tau, \mSigma_\tau),
\end{equation}
 the posterior distribution can be shown to be Gaussian as well. Its mean $\boldsymbol{\mu}_\alpha$ and covariance $\mSigma_\alpha$ are given as
\begin{equation} \label{equ:posterior_coeffs}
\begin{split}
   \bm{\Sigma}_{ \bm{\alpha} } & = \left( \sigma_m^{-1} \mV^T  \mV + \bm{\Sigma}_{\bm{\tau}}^{-1} \right)^{-1} \; \; \textrm{and} \\
   \bm{\mu}_{\bm{\alpha}}^T & = \left( \sigma_m^{-1} \vy^{T} \mV + \bm{\mu}_{\bm{\tau}}^{T}  \bm{\Sigma}_{\bm{\tau}}^{-1} \right)\bm{\Sigma}_{\bm{\alpha}} .
\end{split}
\end{equation}
The approximant $g(\vx)$ is then a GP,
\begin{equation} 
   g\left( \bm{x} \right) \sim \mathcal{N}\left( \bm{\mu}_{g} \left( \boldsymbol{x} \right), \bm{\Sigma}_{g} \left(\boldsymbol{x}, \boldsymbol{x}' \right) \right),
\label{equ:bp}
\end{equation}
with mean and covariance functions
\begin{equation} \label{equ:pred_func}
   \bm{\mu}_{g}\left( \boldsymbol{x} \right) = \vv \left(\boldsymbol{x} \right)^T \bm{\mu}_{\bm{\alpha}} \;  \; \textrm{and} \; \;  \bm{\Sigma}_{g}\left(\vx, \vx'\right) = \vv \left(\vx \right)^{T} \bm{\Sigma}_{ \bm{\alpha} } \vv \left(\vx'\right).
\end{equation}
Here, the vector $\vv(\vx)$ is defined such that $[\vv(\vx)]_j = \phi_j(\vx)$. This process is represented as a graphical model in \Cref{fig:graph1}, where nodes denote random quantities (noting that $\vx$ is also a random quantity distributed according to $\boldsymbol{\rho}$, as described in \Cref{sub:prelim}). 

Note that for $M>N$, in the limit of $\mSigma_\tau\rightarrow \mathbf{0}$, we recover the \emph{least-squares} solution with $\boldsymbol{\mu}$. This corresponds well to the intuitive notion of relying entirely on data, but breaks down when $M<N$.  However, through the use of the Bayesian framework, the situation where $M\leq N$ is handled sensibly: $\mSigma_\tau$ provides a regularising action to the underdetermined problem, and the uncertainty associated with insufficient data is expressed quantitatively and naturally.

\subsection{Designing the coefficient prior}
The specification of the prior $\mathrm{p}(\boldsymbol{\alpha})$ is one aspect of BPC that offers flexibility for the user. In the absence of abundant data about the QoI, the predictive accuracy of the posterior mean polynomial can nevertheless be improved by deliberate design of the prior distribution. Two situations where this is applicable are discussed below.

\subsubsection{Physically informed prior} 
Suppose one has an abundance of data from a distinct but similar quantity $\widetilde{f}(\vx)$ (e.g., a lack of experimental data but sufficient resources to evaluate computational simulations emulating the system). Using the same polynomial basis, one can transfer knowledge from the well-understood $\widetilde{f}$ to improve predictive accuracy of the QoI of which we have a limited understanding. This can be achieved by setting the prior distribution to be centred around the posterior (or frequentist least-squares) coefficients of $\widetilde{f}$. 

\subsubsection{Structural prior}
 Apart from the values of the coefficients, the prior can also influence the structure of the posterior, such as its sparsity. The belief that the QoI depends on the input anisotropically implies that only a small fraction of the coefficients should be non-zero. This knowledge can be leveraged to encourage sparse solutions and improve predictive and generalisation accuracies due to the bias-variance trade-off. In classical PC, sparsity in the fitted model coefficients is promoted by regression methods based on compressed sensing or the least absolute shrinkage and selection operator (LASSO) \cite{tibshirani1996regression}. 

In the Bayesian setting, sparsity can be promoted via design of the prior distribution of the coefficients. Examples of \emph{shrinkage priors} include the spike-and-slab prior \cite{ishwaran2005spike}, the horseshoe prior \cite{carvalho2009handling} and the regularised horseshoe prior \cite{piironen2017sparsity}. The regularised horseshoe prior is specified as
\begin{equation} \label{equ:sparse_prior}
\begin{split}
\widetilde{\lambda_i}, \widetilde{\tau} &\sim \text{HalfCauchy} (1)\\
c^2 &\sim \text{InverseGamma} \left(\frac{\nu}{2}, \frac{\nu}{2}s^2 \right)\\
\lambda_i &= \frac{c\widetilde{\lambda_i}}{\sqrt{c^2 + \tau^2\lambda_i^2}}\\
\alpha_i &\sim \mathcal{N}(0, \tau \lambda_i),
\end{split}
\end{equation}
for each coefficient $i$. The hyperparameters $\nu$ and $s$ are often set empirically, while
\begin{equation}
\tau = \frac{\beta \sqrt{\sigma_m^2}}{(1-\beta) \sqrt{M} },
\end{equation}
where $0<\beta < 1$ controls the degree of sparsity expected in the posterior coefficients; $\sigma_m^2$ is the data variance, and $M$ the number of training points. 

\subsection{On non-Gaussian likelihoods}

When interpreted as a probability distribution function, the likelihood \eqref{eqn:likelihood} can be thought of as a measure of the uncertainty upon fixing observed model parameters. Heuristically, one can assign a Gaussian to characterise this uncertainty. However, in scenarios where one can explicitly state the source of this uncertainty, a Gaussian likelihood may be replaced with a more suitable alternative. One example of this is where epistemic uncertainties are embedded within certain hyperparameters which are involved in simulations but are difficult to ascertain \emph{a priori}. 

In this case, an empirical distribution can be defined with weights on discrete points in the space of hyperparameters, thus forming an approximation of the likelihood function via Monte Carlo. Using this method, a sequential procedure can be used to update the posterior as we draw samples from the empirical distribution, also known as \emph{sequential Monte Carlo}. We leave further investigations on this method as future work.

\subsection{On experimental design}
When forming polynomial approximations of computational models, it is often assumed that the user has control over the formation of the training dataset, especially if a non-intrusive approximation approach is taken. The placement of the input-output pairs can be \emph{designed} by the user. Indeed, the stability of polynomial approximations is strongly influenced by the location of the training inputs. For example, choosing to place the points at the roots of Chebyshev polynomials leads to a much better conditioned polynomial interpolation problem than placing the points uniformly \cite{trefethen2013approximation}. In classical PC and approximations, various deterministic and random strategies have been studied; see \cite{seshadri2019quadrature} for a review.

In the Bayesian setting, a similar question can be asked: where should one place input points to reduce the approximation error and uncertainties? Viewing a Bayesian polynomial as a GP \eqref{equ:bp}, a sequential procedure can be formulated, such that a new input point can be placed at a location where the current posterior variance is largest, representing the largest degree of uncertainty. The challenge in this approach is its generalisation to multiple dimensions, because it involves multivariate polynomial optimisation, a non-convex problem in general. We deem further discussion of this issue beyond the scope of this paper.

\section{Computing and conditioning on moments} \label{sec:moments}
One of the salient advantages of PC is the relatively easy computation of moments and sensitivities. For instance, the mean of the output QoI with respect to the prescribed input uncertainties is given by the first coefficient of the polynomial approximation; the variance is given by the sum of the squares of the remaining coefficients (see page 210 in \cite{smith2013uncertainty}). For Bayesian polynomials, the posterior uncertainties associated with the coefficients are analogously propagated to the moments. For example, the output (spatial) mean is given by
\begin{equation}
   \begin{split}
   \mathbb{E}_{\vx}\left[g\right] = & \int_{\mathcal{D}} g\left(\vx \right) \bm{\rho}\left( \vx \right) d \vx \\
   = & \int_{\mathcal{D}} \vv(\vx)^T \boldsymbol{\alpha}~ \bm{\rho}\left( \vx\right) d \vx\\
   = & \left(\int_{\mathcal{D}} \vv(\vx) ~ \bm{\rho}\left( \vx\right) d \vx\right)^T \boldsymbol{\alpha}
   \end{split}
   \label{equ:mean}
\end{equation}
Letting $\vv_{int} := \left(\int_{\mathcal{D}} \vv(\vx) ~ \bm{\rho}\left( \vx\right) d \vx\right)$ and assuming Gaussianity of the coefficient posterior, we have
\begin{equation}
\begin{split}
   \mathbb{E}_{\vx}\left[g\right] \sim & \mathcal{N}(\vv_{int}^T \boldsymbol{\mu}_{\boldsymbol{\alpha}}, \vv_{int}^T \boldsymbol{\Sigma}_{\alpha} \vv_{int})\\
  =  & \; \mathcal{N}\left( [\bm{\mu}_{\bm{\alpha}}]_{1},  [\bm{\Sigma}_{\bm{\alpha}}]_{1,1} \right)
   \end{split}
\end{equation}
because $\vv_{int} = [1, 0, ..., 0]^T$ owing to the orthogonality of the basis polynomials. In a similar manner, we write the variance as
\begin{equation}
   \begin{split}
   \textrm{Var}\left[g\right] =  &\int_{\mathcal{D}} g^2\left( \bm{x} \right) \rho\left( \bm{x} \right) d\vx - \left( \mathbb{E}\left[g\right] \right)^2 \\
   &= \boldsymbol{\alpha}^T \left(\int_{\mathcal{D}}  \vv(\vx) \vv(\vx)^T \rho(\vx)d\vx\right) \boldsymbol{\alpha} - \left( \boldsymbol{\alpha}^T \int_{\mathcal{D}}  \vv(\vx) ~ \bm{\rho}\left( \boldsymbol{x} \right) d \bm{x}\right)^2 \\
   &= \boldsymbol{\alpha}^T \left(\mSigma_{int} - \vv_{int}\vv_{int}^T \right) \boldsymbol{\alpha} \\
   &= \boldsymbol{\alpha}^T 
   \begin{bmatrix}
   0 & & & \\
    & 1 & & \\
    & & \ddots & \\
    & & & 1
\end{bmatrix}    \boldsymbol{\alpha} \\
&= \sum_{i=2}^N \alpha_i^2,
   \end{split}
   \label{equ:var}
\end{equation}
where on the third line, $\mSigma_{int} = \int_{\mathcal{D}}  \vv(\vx) \vv(\vx)^T \rho(\vx)d\vx = \mI$ owing to the orthogonality of the basis polynomials. This can be expressed as a quadratic form of $\boldsymbol{\alpha}$, which has a generalised chi-squared distribution (see, e.g., \cite{das2021method}) or, equivalently, a weighted sum of chi-squared and Gaussian distributions (see Appendix \ref{sec:appen}).

Expressions for conditional variances and the related Sobol' indices can also be derived. First, we introduce the function $\mathcal{S}$ that \emph{subselects} rows of the multi-index set $\mathcal{B}$ that have an order of $p_\Delta$ along direction $\Delta$,
\begin{equation}
\widetilde{\mathcal{B}} = \mathcal{S}\left( \mathcal{B}, p_\Delta, \Delta \right)
\end{equation}
where $1 \leq \Delta \leq d$. 
\begin{equation}
  \sigma_S\left[g\right] =  \frac{\sum_{i\in \widetilde{\mathcal{B}}} \alpha_i^2}{\textrm{Var}\left[g\right] },
   \label{equ:sobol}
\end{equation}
which is a ratio of mutually correlated generalised chi-squared variables. Further details on the characterisation of these quantities are omitted for brevity, since one can straightforwardly sample from these distributions using samples of the coefficients and pushing them through \eqref{equ:sobol}.

\subsection{Conditioning on linear operators} \label{sub:cond_moments}

In some engineering applications, it is often easier to estimate integral-, differential-, or more generally linear-operators of related QoIs than the required QoI itself. In line with earlier remarks, this data need not arise from simulations---it can stem from experimental results or even expert knowledge. If probabilistic descriptions of such data are known, then they can be used to constrain the space of posterior polynomial distributions. In this subsection, we assume that the coefficient prior is Gaussian such that the Bayesian polynomial is a GP, so that analytical expressions can be derived. However, we stress that the following technique is general and does not require assumption of Gaussianity.

Suppose that we have a linear operator $\mathcal{L}: \mathbb{R} \rightarrow \mathbb{R}^L$ operating on scalar output values. We formalise this idea by considering the joint distribution on a collection of points $\mX$ in the input domain 
\begin{equation} \label{equ:cond_distr}
\begin{split}
\left(\begin{array}{c}
g \left( \bm{\mX} \right) \\
\mathcal{L}\left\{ g \left( \mX \right)  \right\}
\end{array}\right) &= \mathcal{N}\left(\left[
\begin{array}{c}
\bm{\mu}_{g} \left( \mX \right) \\
\mathcal{L}\left\{ \bm{\mu}_{g} \left( \mX \right)  \right\}
\end{array}
\right],\left[
\begin{array}{cc}
\bm{\Sigma}_{g}\left(\mX,\mX \right) & \mathcal{L}'\left\{ \bm{\Sigma}_{g}\left(\mX, \mX' \right)\right\} \\
\mathcal{L}'\left\{ \bm{\Sigma}_{g}\left(\mX,\mX'\right)\right\} ^{T} & \mathcal{L}' \left\{ \mathcal{L}'' \left\{ \bm{\Sigma}_{g}\left(\mX',\mX''\right)\right\} \right\}
\end{array}\right]\right) \\
&= \mathcal{N}\left(\left[
\begin{array}{c}
\bm{\mu}_{1} \\
\bm{\mu}_{2}
\end{array}
\right],\left[
\begin{array}{cc}
\mSigma_{11} & \mSigma_{12}\\
\mSigma_{12}^T & \mSigma_{22}
\end{array}\right]\right)
\end{split}
\end{equation}
where the linearity of $\mathcal{L}$ has been used to exchange the order of operation of expectations with the linear operator. The notations $\mathcal{L}'$ and $\mathcal{L}''$ indicate that the linear operators are applied on $\mX'$ and $\mX''$ (identical copies of $\mX$) respectively. 

From this equation, we can write the conditional distribution of the polynomial conditioned on a value of the linear functional $g | (\mathcal{L}\left\{ g \right\} = \va)$ for some $\va \in \mathbb{R}^L$ using standard Gaussian identities. This is given by
\begin{equation} 
g(\mX) \;|\; (\mathcal{L}\left\{ g\left( \bm{\mX} \right)\right\} = \va) \sim \mathcal{N}\left(\bm{\mu}_{c}\left( \mX, \va \right),\mSigma_{c}\left(\mX,\mX' \right)\right),
\end{equation}
where
\begin{equation}\label{equ:pred_condmean}
\bm{\mu}_{c}\left(\mX, \va\right)  = \bm{\mu}_{1}+ \mSigma_{12}^{T} \mSigma_{22}^{-1} \left( \va - \bm{\mu}_2 \right)
\end{equation}
and
\begin{equation}
\mSigma_{c}\left(\mX,\mX'\right) =\mSigma_{11} - \mSigma_{12} ^{T} \mSigma_{22} ^{-1} \mSigma_{12}.
\end{equation}

Suppose further that we have information on the linear operators up to a certain confidence. That is, $\va \sim \mathcal{N}(\bm{\mu_{\va}}, \mSigma_{\va})$ for certain known $\bm{\mu_{\va}}$ and $\mSigma_a$. Then, the mean prediction is
\begin{equation} 
\mathbb{E} \left[\mathbb{E}\left[g(\mX) \;|\; \mathcal{L}\left\{ g\left( \bm{\mX} \right)\right\}\right] \right] = \bm{\mu}_{1} + \mSigma_{12}^{T} \mSigma_{22}^{-1} \left( \bm{\mu_{\va}} - \bm{\mu}_2 \right).
\end{equation}
Note the effect of introducing conditioning on the linear functional: the covariance of the prediction is a correction to the posterior prediction given by the second term above. The covariance of the prediction is 
\begin{equation}
\begin{split}
\text{Cov} \left[ g(\mX) \right] &= \mathbb{E} \left[\text{Cov}\left[g(\mX) \;|\; \mathcal{L}\left\{ g\left( \bm{\mX} \right)\right\} \right] \right] + \text{Cov} \left[\mathbb{E}\left[g(\mX) \;|\; \mathcal{L}\left\{ g\left( \bm{\mX} \right)\right\}\right]\right]\\
&= \mSigma_{c}  + \mSigma_{12}^{T} \mSigma_{22}^{-1} \mSigma_{\va}  \mSigma_{22}^{-T} \mSigma_{12}.
\end{split}
\end{equation}
\section{Coregional models} \label{sec:coregional}
When constructing models for multiple objectives that are functions of the same set of input parameters, one can improve predictive performance by exploiting correlations between different outputs by leveraging heuristics under \emph{transfer learning}. In this section, we take clues from a transfer learning method for GP regression, known as the \emph{intrinsic model of coregionalisation} \cite{journel1978mining,cressie1993statistics,mauricio2011computationally}, an instance of \emph{cokriging}, to construct multi-task Bayesian polynomial models that are able to transfer information between the modelled outputs to improve predictive accuracy over individually trained models.

\subsection{An alternative Gaussian process formulation}
Before proceeding, we introduce an alternative, but equivalent, formulation of BPC based on GP regression. Recall from \eqref{equ:bp} that a Bayesian polynomial with Gaussian distributed coefficients can be regarded as a GP. In fact, this idea can be extended: the mean prediction of a GP with an orthogonal polynomial kernel can be regarded as an evaluation of an orthogonal polynomial expansion. To show this, define a GP,
\begin{equation}
f \sim GP\left(0, k(\vx, \vx_\ast)\right),
\end{equation}
where the \emph{covariance function} $k(\vx, \vx_\ast)$ is defined with an orthogonal polynomial kernel,
\begin{equation}
k(\vx, \vx_\ast) = \vv(\vx)^T \mSigma \vv(\vx_\ast),
\end{equation}
where $\vv(\vx)$ is the evaluation of the polynomial basis functions at $\vx$, as defined in \eqref{equ:pred_func}, and $\mSigma$ is a positive definite matrix. Define the kernel function on two sets of input points $\mX = [\vx_1, \vx_2,..., \vx_M]$ and $\mX_\ast = [\vx_{\ast 1}, \vx_{\ast 2},..., \vx_{\ast N}]$, yielding
\begin{equation}
K(\mX, \mX_\ast) = \begin{bmatrix}
k(\vx_1, \vx_{\ast 1}) & k(\vx_1, \vx_{\ast 2}) & \cdots & k(\vx_1, \vx_{\ast N}) \\
k(\vx_2, \vx_{\ast 1}) & k(\vx_2, \vx_{\ast 2}) & \cdots & k(\vx_2, \vx_{\ast N}) \\
\vdots & \vdots & \ddots & \vdots \\
k(\vx_M, \vx_{\ast 1}) & k(\vx_M, \vx_{\ast 2}) & \cdots & k(\vx_M, \vx_{\ast N})
\end{bmatrix}.
\end{equation}
In GP regression, the posterior prediction on a set of test points $\mX_\ast$ given noisy output values $\vy$ (with noise variance $\sigma_m^2$) on training points $\mX$ is given by
\begin{equation} \label{eqn:coreg_gp_post}
\begin{split}
f_{pred}(\mX_\ast) &= K(\mX_\ast, \mX) \left(K(\mX, \mX) + \sigma_m\mI\right)^{-1} \vy \\
&= \mV(\mX_\ast) \underbrace{\mSigma \mV(\mX)^T \left(K(\mX, \mX) + \sigma_m\mI\right)^{-1} \vy }_{\boldsymbol{\alpha}_{post}},
\end{split}
\end{equation}
which can be expressed as the product between the polynomial design matrix on the test points $\mV(\mX_\ast)$ and a vector $\boldsymbol{\alpha}_{post}$ independent of $\mX_\ast$, which can be considered the posterior coefficients. Although the kernel in this situation is finite dimensional, introducing this GP formulation permits us to construct correlations between predictions at different inputs (and potentially from different models, as will be explained next) instead of the less intuitive correlations drawn among polynomial coefficients. 

\subsection{Bayesian polynomial chaos as a coregional Gaussian process}
Using the formulation described above, correlations between multiple outputs can be enforced by incorporating them in the covariance function. In detail, the following covariance relation is stipulated for two outputs $y_i$ and $y_j$ evaluated at two possibly different input locations $\vx_i$ and $\vx_j$:
\begin{equation}
\text{cov}(y_i(\vx_i), y_j(\vx_j)~|~\mA, \mB, \sigma_m^2) = \vv(\vx_1)^T \sqrt{\mSigma_i \mSigma_j} \vv(\vx_j) \times B_{ij} + \sigma_m \delta_{ij},
\end{equation}
where we specify that $\mSigma_i$ is a diagonal matrix with positive entries, $\mSigma_i = \text{diag}(\va_i) = \text{diag}(\alpha_{i1}, \alpha_{i2}, ..., \alpha_{iN})$ representing the variance of the coefficients of $i$. Here, $\va_i$ denotes the $i$-th column of $\mA$, which collects the coefficient variance parameters $\alpha_{ij}$ such that the $(i, j)$ entry of $\mA$ is $\alpha_{ij}$. Also, $B_{ij}$ is the $(i, j)$ entry of a matrix $\mB$, modulating the correlation between the outputs. When $i=j$, the Kronecker delta $\delta_{ij} = 1$ which adds the measurement noise variance $\sigma_m$ to the covariance; otherwise, this variance is not added. Thus, one can write down the covariance matrix of the stacked outputs without noise,
\begin{equation}
\vf(\mathcal{X}) = \left[ f_1(\mX_1)^T \; f_2(\mX_1)^T \; ... \; f_O(\mX_1)^T\right]^T,
\end{equation}
where $f_i(\mX_i) = [f_i(\vx_{i1}), f_i(\vx_{i2}), ..., f_i(\vx_{iM})]$ and $\mathcal{X} = [\mX_1, \mX_2, ..., \mX_O]$, as 
\begin{equation}
\begin{split}
&\text{cov}(\vf ~|~\mA, \mB, \sigma_m^2) := \mK(\mathcal{X}, \mathcal{X}) \\
&= \begin{bmatrix}
\mV(\mX_1) \mSigma_1 \mV(\mX_1)^T B_{11} & \mV(\mX_1) \sqrt{\mSigma_1\mSigma_2} \mV(\mX_2)^T B_{12} & \cdots & \mV(\mX_1) \sqrt{\mSigma_1\mSigma_O} \mV(\mX_O)^T B_{1O} \\
\mV(\mX_2) \sqrt{\mSigma_2\mSigma_1} \mV(\mX_1)^T B_{21} & \mV(\mX_2) \mSigma_2 \mV(\mX_2)^T B_{22} & \cdots & \mV(\mX_2) \sqrt{\mSigma_2\mSigma_O} \mV(\mX_O)^T B_{2O} \\
\vdots & \vdots & \ddots & \vdots \\
\mV(\mX_O) \sqrt{\mSigma_O\mSigma_1} \mV(\mX_1)^T B_{O1} & \mV(\mX_O) \sqrt{\mSigma_O\mSigma_2} \mV(\mX_2)^T B_{O2} & \cdots & \mV(\mX_O) \mSigma_O \mV(\mX_O)^T B_{OO}
\end{bmatrix}.
\end{split}
\end{equation}
From this, defining 
\begin{equation}
\mK_{11} = K(\mathcal{X}, \mathcal{X}), \; \mK_{21} = K(\mathcal{X}_\ast, \; \mathcal{X}) = \mK_{12}^T, \; \mK_{22} = K(\mathcal{X}_\ast, \mathcal{X}_\ast), 
\end{equation}
with $\mathcal{X}$ and $\mathcal{X}_\ast$ being the training and testing points for all $O$ outputs respectively. One can use the standard expression for posterior prediction:
\begin{equation} \label{eqn:coreg_post_pred}
\vf(\mathcal{X}_\ast)~|~ \vy(\mathcal{X}), \mA, \mB, \sigma_m^2 \sim \mathcal{N}(\mK_{21} (\mK_{11} + \sigma_m \mI)^{-1} \vy, \mK_{22} - \mK_{21}(\mK_{11} + \sigma_m^2 \mI)^{-1} \mK_{12}).
\end{equation}
However, the quantities $\mA, \mB, \sigma_m^2$ are unknown \emph{a priori}. In our approach, we let $\mA$ and $\mB$ be random variables, with prior distributions 
\begin{equation} \label{eqn:coreg_hyp}
\alpha_{ij} \sim \mathcal{N}(0, 1),\quad \mB = \mW\mW^T + \text{diag}(\kappa),\quad \mW \sim \mathcal{N}(0, \mI),\quad \kappa \sim \text{HalfNormal}(1)
\end{equation}
where $\mW \in \mathbb{R}^{O\times 1}$ and $\kappa$ contains positive entries. The full distribution of the outputs is thus non-Gaussian; to deduce the mean and variance of the full posterior distribution of the outputs, one can use the laws of iterated expectation and total covariance, where expectations can be taken by sample means using samples obtained from the posterior distribution. That is, samples of $\boldsymbol{\alpha}_{post}$ as defined in \eqref{eqn:coreg_gp_post} are obtained from Markov chain Monte Carlo by combining posterior samples of the hyperparameters in \eqref{eqn:coreg_hyp}. The predictive mean and covariance of a set of points $\mathcal{X}_\ast$ is then
\begin{equation}
\begin{split}
\mathbb{E}[\vf(\mathcal{X}_\ast)] &= \mathbb{E}[\mathbb{E}[\vf(\mathcal{X}_\ast)~|~\boldsymbol{\alpha}_{post}]]\\
\text{cov}(\vf(\mathcal{X}_\ast)) &= \mathbb{E}[\text{cov}(\vf(\mathcal{X}_\ast)~|~\boldsymbol{\alpha}_{post})] + \text{cov}(\mathbb{E}[\vf(\mathcal{X}_\ast)~|~\boldsymbol{\alpha}_{post}]),
\end{split}
\end{equation}
which can be evaluated using samples of $\boldsymbol{\alpha}_{post}$ substituted into \eqref{eqn:coreg_post_pred}.

\section{Numerical results} \label{sec:examples}
In this section, we demonstrate the BPC framework by applying the various methods developed in the prior sections to analytical and real-world datasets. The focus in this section is not merely to demonstrate that the Bayesian formulation allows additional uncertainties, associated with the quantity of data and the quality of the model from which the data is sourced, to be quantified; we also show that various techniques employed in the solution of the posterior distribution can help improve the predictive accuracy.

\subsection{Conditioning on the spatial mean}
As established in \Cref{sub:cond_moments}, by introducing approximate knowledge of the value of linear functionals of the output, one can introduce a correction term in the posterior prediction. An example of a linear operator on the output is the output QoI mean,
\begin{equation}
\mathcal{L} \{g(\vx)\} = \int_\mathcal{D} g(\vx) ~ \rho(\vx) ~ d\vx \in \mathbb{R}.
\end{equation}
Using the orthogonality of the basis,
\begin{equation}
\begin{split}
\mathcal{L}' \{ \mSigma_g (\vx, \vx') \} &= \mathcal{L}' \{ \vv(\vx)^T \mSigma_{\alpha} \vv(\vx') \} = \mathcal{L}' \left\{\sum_{i, j = 1}^{N} [\Sigma_{\alpha}]_{ij} \phi_i(\vx) \phi_j(\vx') \right\} \\
&=  \sum_{i, j = 1}^{N} [\Sigma_{\alpha}]_{ij} \phi_i(\vx) \delta_{j1} \\
&= \vv(\vx) ^T [\mSigma_{\alpha}]_{1}
\end{split}
\end{equation}
\begin{equation}
\begin{split}
\mathcal{L}' \left\{ \mathcal{L}'' \left\{ \bm{\Sigma}_{g}\left(\vx',\vx''\right)\right\} \right\} &= \mathcal{L}' \left\{ \mathcal{L}'' \left\{ \sum_{i, j = 1}^{N} [\Sigma_{\alpha}]_{ij} \phi_i(\vx') \phi_j(\vx'') \right\} \right\} \\
&= \sum_{i, j = 1}^{N} [\Sigma_{\alpha}]_{ij}\delta_{i1} \delta_{j1} \\
&= [\mSigma_{\alpha}]_{11}.
\end{split}
\end{equation}
Here $[\mSigma_{\alpha}]_{1}$ denotes the first column of $\mSigma_{\alpha}$ and $[\mSigma_{\alpha}]_{11}$ is the top left entry of $\mSigma_{\alpha}$. The component quantities in \eqref{equ:cond_distr} can be expressed as follows:
\begin{equation}
\mu_2 = \left[\bm{\mu}_{\bm{\alpha}} \right]_1
\end{equation}
\begin{equation}
\mSigma_{12} = \mV (\mX) [\mSigma_{\alpha}]_{1}
\end{equation}
\begin{equation}
\Sigma_{22} = [\mSigma_{\alpha}]_{11}
\end{equation}
where $\mV (\mX) \in \mathbb{R}^{M\times N}$ is the Vandermonde-type matrix, and $\left[\bm{\mu}_{\bm{\alpha}} \right]_1$ is the first entry of $\bm{\mu}_{\bm{\alpha}}$.

To demonstrate the correctional effect of conditioning on the spatial mean, consider the following numerical example pertaining to the modelling of the von Karman Institute (VKI) LS89 axial jet engine turbine blade. In Arts et al. \cite{arts1992aero}, an experimental campaign was carried out on a range of exit Mach numbers $Ma$, exit Reynolds numbers $Re$, and freestream turbulence intensities $Ti$. These three quantities constitute the input parameters in this test case. To these we assign independent uniform distributions bounded as follows: 
\begin{equation}
\begin{split}
0.7 \leq ~&Ma \leq 1.1,\\
5\times 10^5 \leq~ &Re \leq 2\times 10^6,\\
1 \leq~ &Ti \leq 6.
\end{split}
\end{equation}
The total wall heat flux $H_{int}$ is chosen as the QoI as a function of $\vx = [Ma, Re, Ti]$. Drawing from the tabulated results in \cite{arts1992aero}, a set of experimental results consisting of 21 input-output pairs is obtained. From this set, we randomly select 15 training data points and assign the rest as test data. 

\begin{figure}
\begin{subfigure}[b]{0.34\textwidth}
\centering
\includegraphics[width=\textwidth]{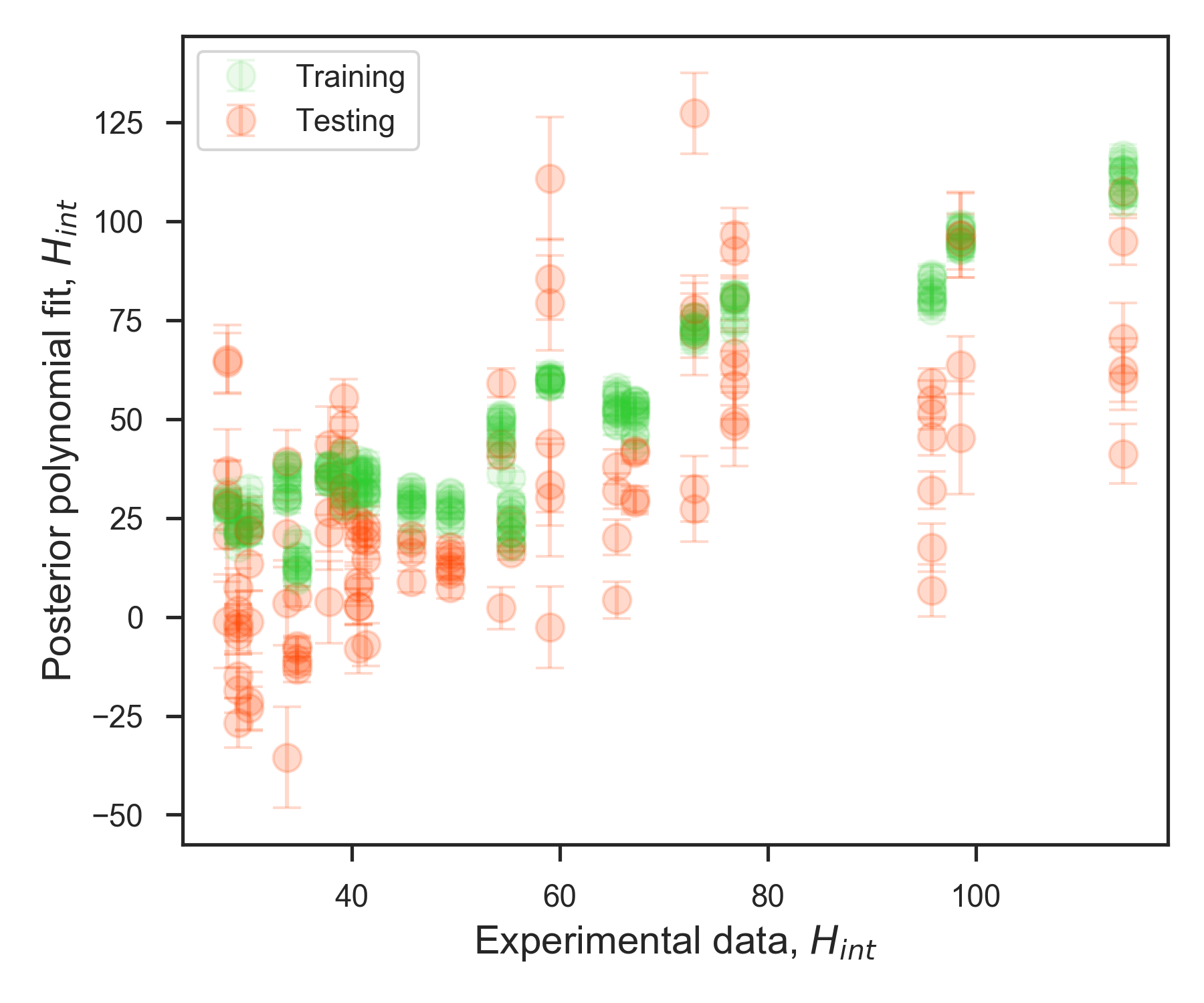}
\caption{No conditioning}
\end{subfigure}
\begin{subfigure}[b]{0.34\textwidth}
\centering
\includegraphics[width=\textwidth]{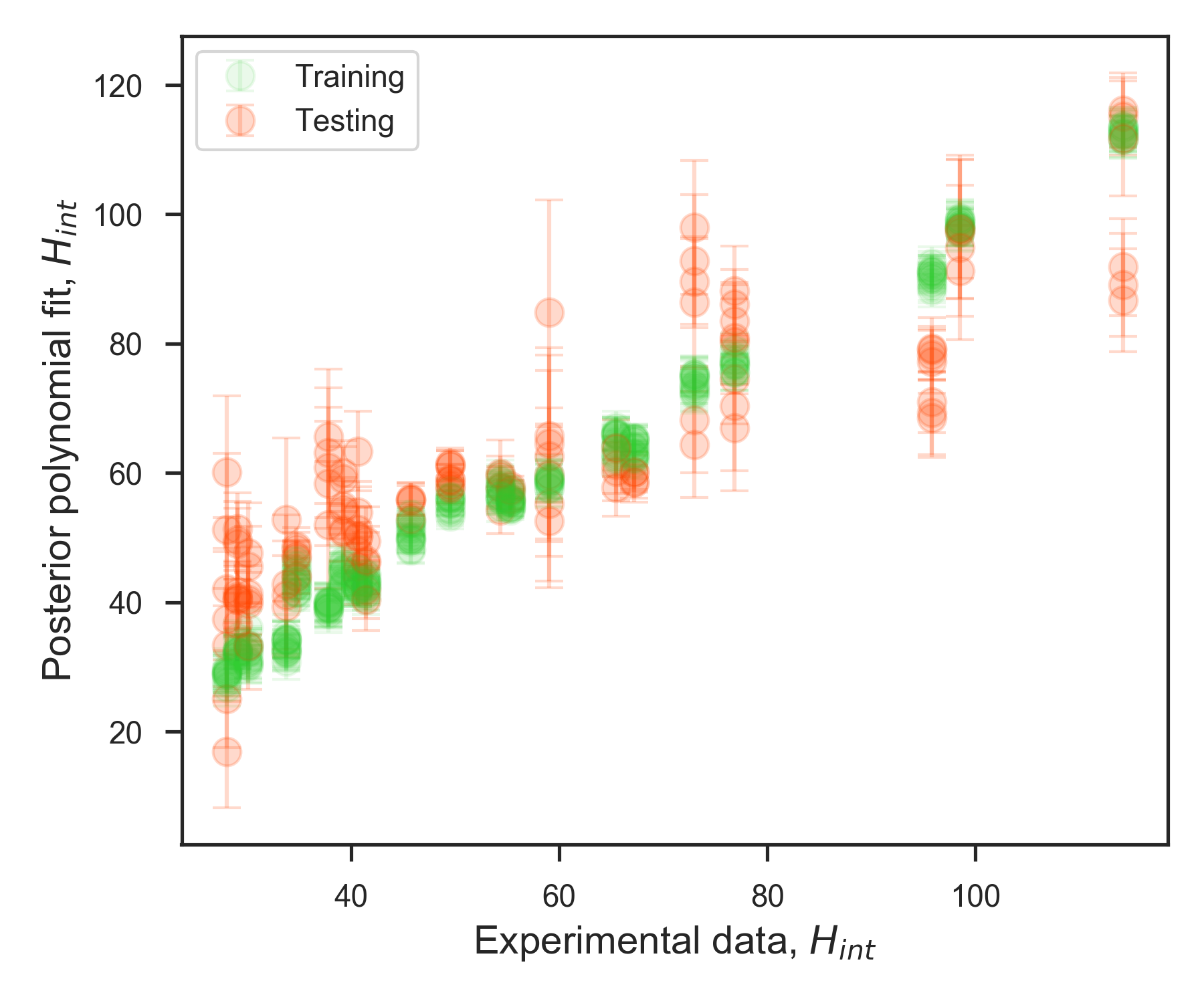}
\caption{With conditioning}
\end{subfigure}
\begin{subfigure}[b]{0.3\textwidth}
\centering
\includegraphics[width=\textwidth]{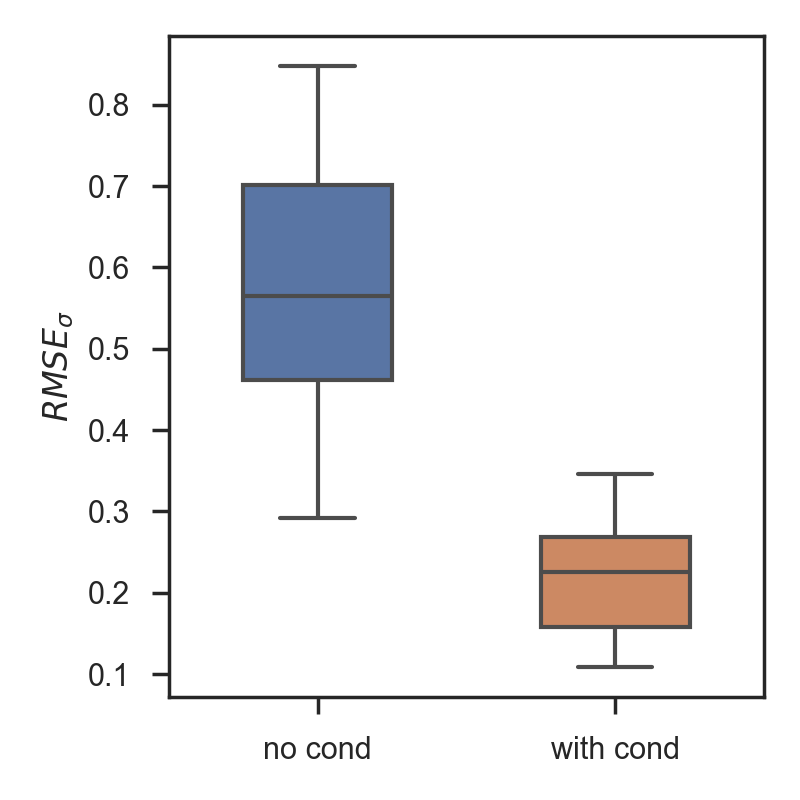}
\caption{RMSE comparison}
\label{fig:condmean_rmse}
\end{subfigure}
\caption{Fitting the VKI data with and without conditioning on the mean.}
\label{fig:condmean_fits}
\end{figure}

We employ a Bayesian polynomial model to predict the experimental values from the three input variables. The polynomial index set is set to be an isotropic total order grid of maximum order 3 (implying that the degrees of polynomials used in each dimension sum up to a maximum of 3), amounting to 20 coefficients. We employ Legendre orthonormal polynomials as the basis $\phi$. In the absence of knowledge about the output, the prior for the coefficients is left as a generic uninformative prior, with $\mathrm{p} (\boldsymbol{\alpha}) \sim \mathcal{N} (\mathbf{0}, \mI)$. The noise variance (see \eqref{eqn:noise_var}) is set at $\sigma_m = 3.0$, a small fraction of the overall output variance. Following \eqref{equ:posterior_coeffs}, the posterior coefficients are Gaussian variables, for which we compute mean and covariance. Then, according to \eqref{equ:pred_func}, the mean function and standard deviation at the test points are evaluated, and the predictions based on the mean function are compared with the experimental data. We repeat this exercise, but with the posterior mean function modified by conditioning on the mean of the experimental data according to \eqref{equ:pred_condmean}. \Cref{fig:condmean_fits} shows the resultant scatter plots from 20 random selections of training data, overlaid in one graph. The predictive accuracy is clearly improved by the information provided by the conditioning on the mean. \Cref{fig:condmean_rmse} further articulates this with the normalised test root-mean-squared error (RMSE), defined as
\begin{equation}
\text{RMSE}_\sigma = \frac{1}{M_{test}\sigma_{out}^{1/2}} \sqrt{\sum_{m=1}^{M_{test} } (y_{m, test} - \bar{y}(\vx^{(m)}) )^2},
\end{equation}
where $\bar{y}(\vx^{(m)}) = \vv(\vx^{(m)}_{test})^T \boldsymbol{\mu}_\alpha$ is the posterior mean prediction at the $m$-th test input $\vx^{(m)}_{test}$.

\subsection{Physically informed priors}
\begin{figure}
\centering
\includegraphics[width=0.7\textwidth]{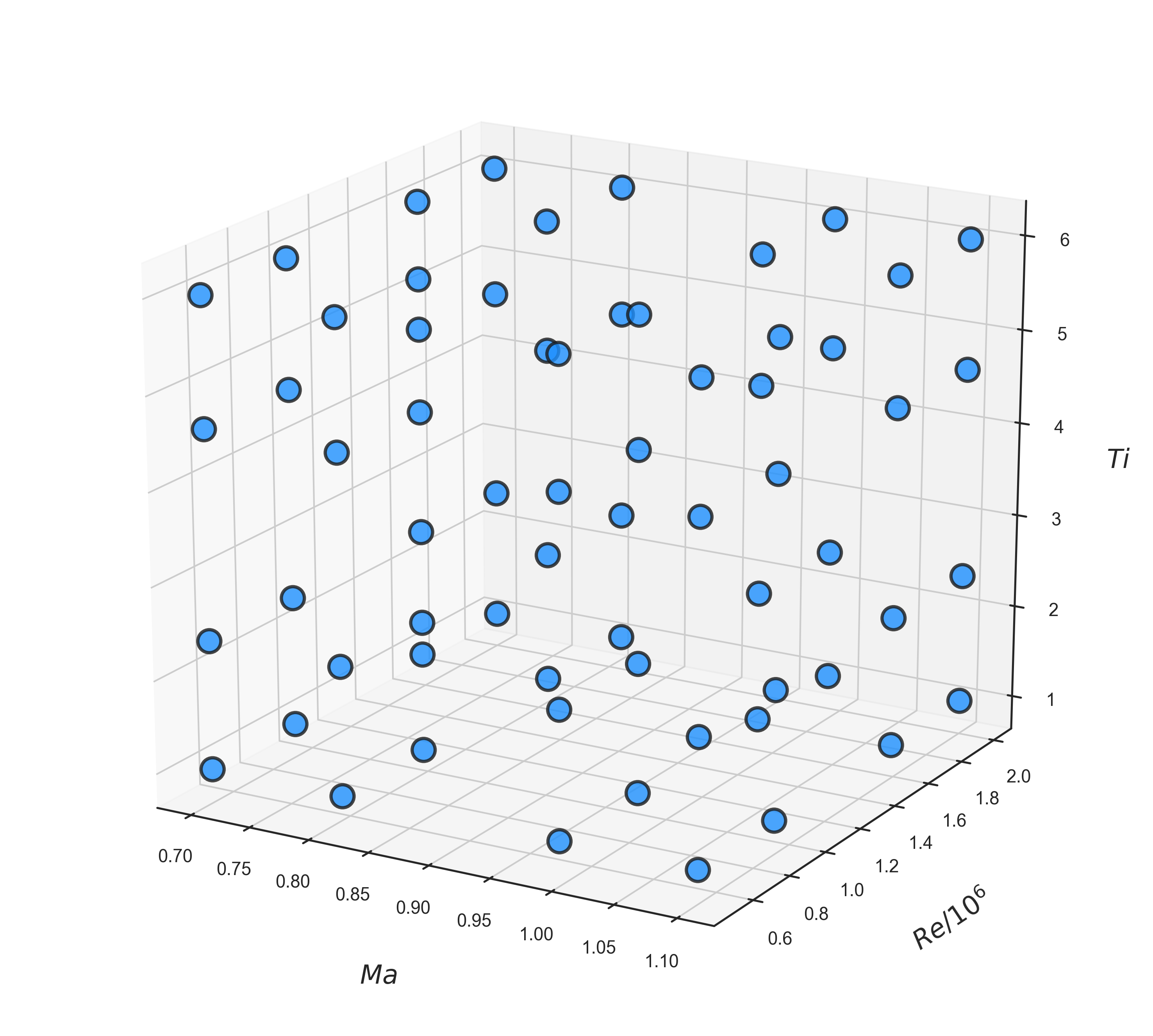}
\caption{Gauss-Lobatto input points where $H_{int}$ is evaluated using CFD. }
\label{fig:gausslobatto}
\end{figure}

Consider again the case of the VKI LS89 turbine. The availability of experimental data is limited, but computational fluid dynamics (CFD) simulations of the same turbine can be used  to mimic the relevant physical behaviour. The open-source CFD solver SU2 \cite{economon_su2:_2016} is used to calculate $H_{int}$ on a grid of input points consisting of 64 Gauss-Lobatto points on a tensor grid with maximum degree 3 on each axis (see \Cref{fig:gausslobatto}). The Reynolds-averaged Navier-Stokes (RANS) equations are solved with Spalart-Allmaras turbulence closure. No turbulence transition model is used and all boundary layers are assumed to be turbulent. 

Owing to the approximations made by the CFD calculations, one cannot expect the data obtained from simulations to completely replace experimental data. Despite this, the dependence of the CFD output on the inputs can be expected to follow a similar pattern to the experimental data. To leverage this similarity, a multivariate polynomial of the three input parameters mentioned above is fitted on a total order grid polynomial basis of maximum degree 3. The fitted coefficients $\bm{\mu}_{CFD}$ can then be used as the prior mean coefficients of the BPC model for the experimental output.

\begin{figure}
\begin{subfigure}[b]{0.34\textwidth}
\centering
\includegraphics[width=\textwidth]{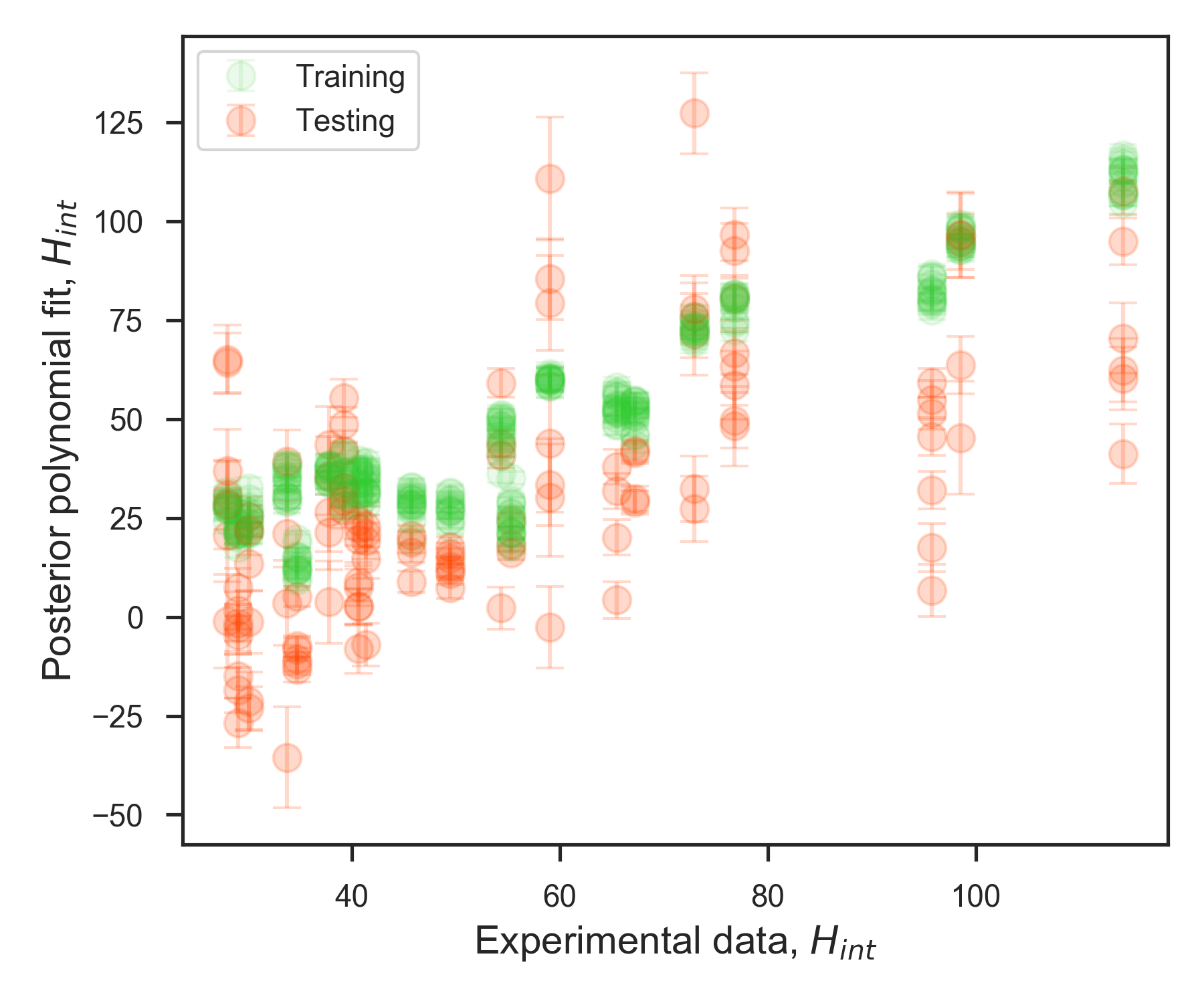}
\caption{Zero prior}
\end{subfigure}
\begin{subfigure}[b]{0.34\textwidth}
\centering
\includegraphics[width=\textwidth]{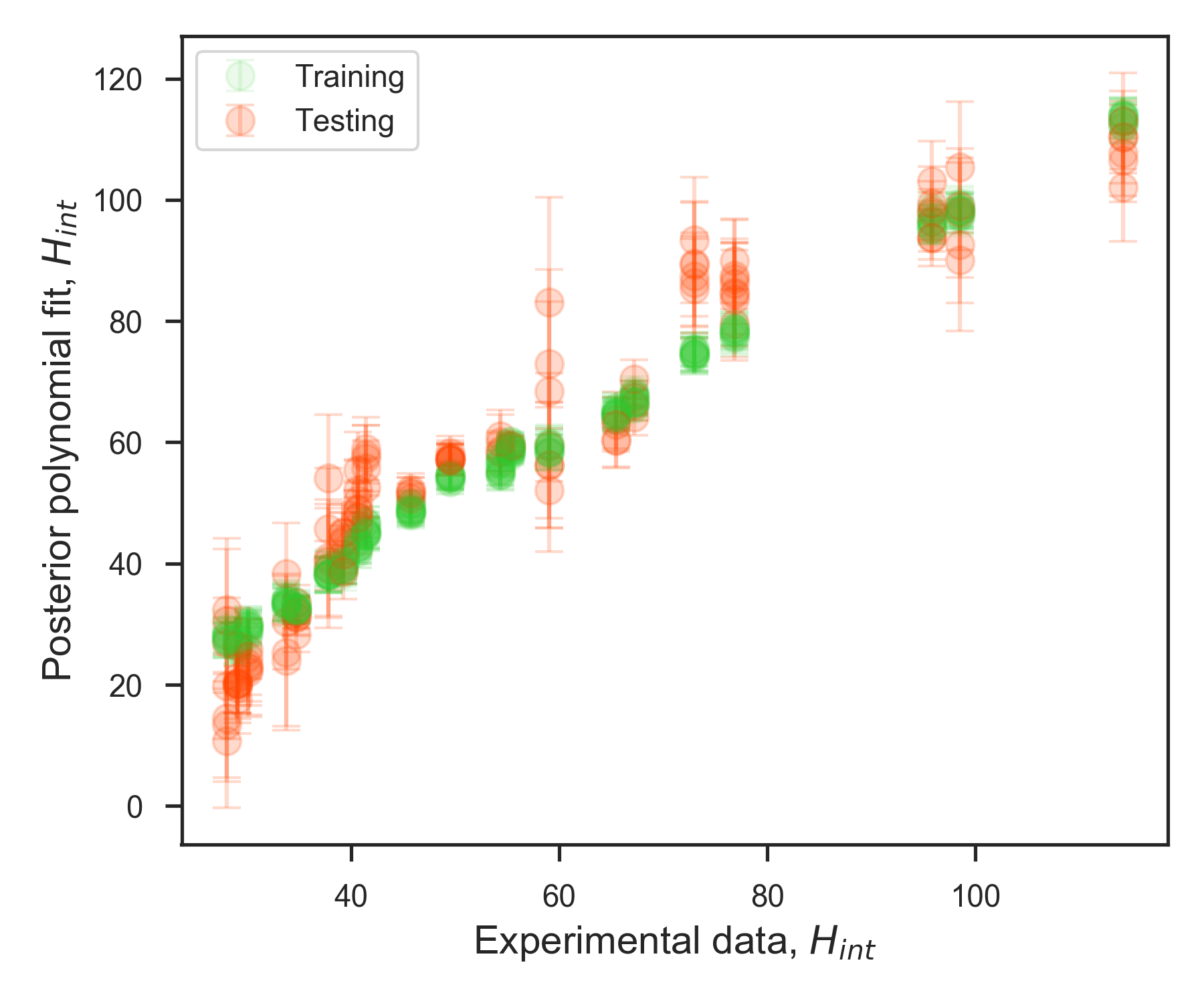}
\caption{CFD prior}
\end{subfigure}
\begin{subfigure}[b]{0.3\textwidth}
\centering
\includegraphics[width=\textwidth]{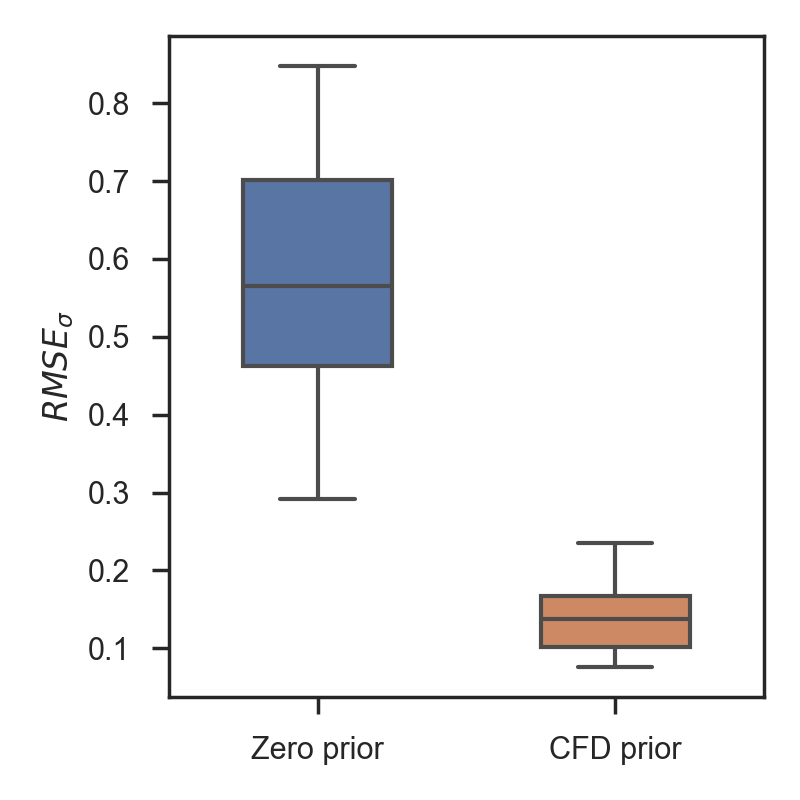}
\caption{RMSE comparison}
\label{fig:vki_rmse}
\end{subfigure}
\caption{Fitting the VKI data with and without CFD prior.}
\label{fig:vki_fits}
\end{figure}

For this example, the prior distribution of coefficients is set as $\mathrm{p}(\boldsymbol{\alpha}) = \mathcal{N}(\boldsymbol{\mu}_p, \mI)$ with data variance $\sigma_m = 3.0$---the same as the previous example except for the prior mean coefficients $\boldsymbol{\mu}_p$ which we compare between two cases,
\begin{itemize}
\item $\boldsymbol{\mu}_p = \mathbf{0}$, and
\item $\boldsymbol{\mu}_p = \bm{\mu}_{CFD}$.
\end{itemize}
The polynomial is constructed over a three-dimensional isotropic total order Legendre basis of maximum order 3 ($\boldsymbol{\alpha} \in \mathbb{R}^{20}$, as before). We use 15 points randomly selected from a pool of 21 points from the experimental data to fit the model over 20 trials, using the remaining points as testing data. The results for the trials are overlaid and shown in \Cref{fig:vki_fits}, comparing the cases of $\boldsymbol{\mu}_p = \mathbf{0}$ and $\boldsymbol{\mu}_p = \bm{\mu}_{CFD}$. It is clear that using the CFD prior results in better predictive accuracy over testing data. \Cref{fig:vki_rmse} summarises this through a boxplot of the test RMSE normalised by the output standard deviation. This plot shows that the RMSE is greatly reduced by using the CFD prior, demonstrating that information can be transferred from the CFD-inferred coefficients to improve predictive accuracy.

\subsection{Structurally informed priors} \label{sub:ex_structural}
In this example, we consider a different dataset consisting of three fan blade geometries, labelled A, B and C. Blades A and C are high-speed fan blades, while blade B is a low-speed fan blade. Each of these blades is parameterised by the same 25 geometry parameters that define the airfoil profile at five spanwise locations. The Rolls-Royce code PADRAM is used to generate the computational meshes, on which the RANS equations are solved using the HYDRA suite to obtain non-dimensionalised fan efficiency values for each geometry \cite{crumpton1998unstructured,lapworth2004hydra}. Further details of the CFD setup can be found in \cite{seshadri2020supporting}. The training datasets for blades A, B and C consist of 548, 381 and 547 input-output points respectively, and are available at \url{https://github.com/psesh/turbodata}.

\begin{figure}
\centering
\includegraphics[width=0.5\textwidth]{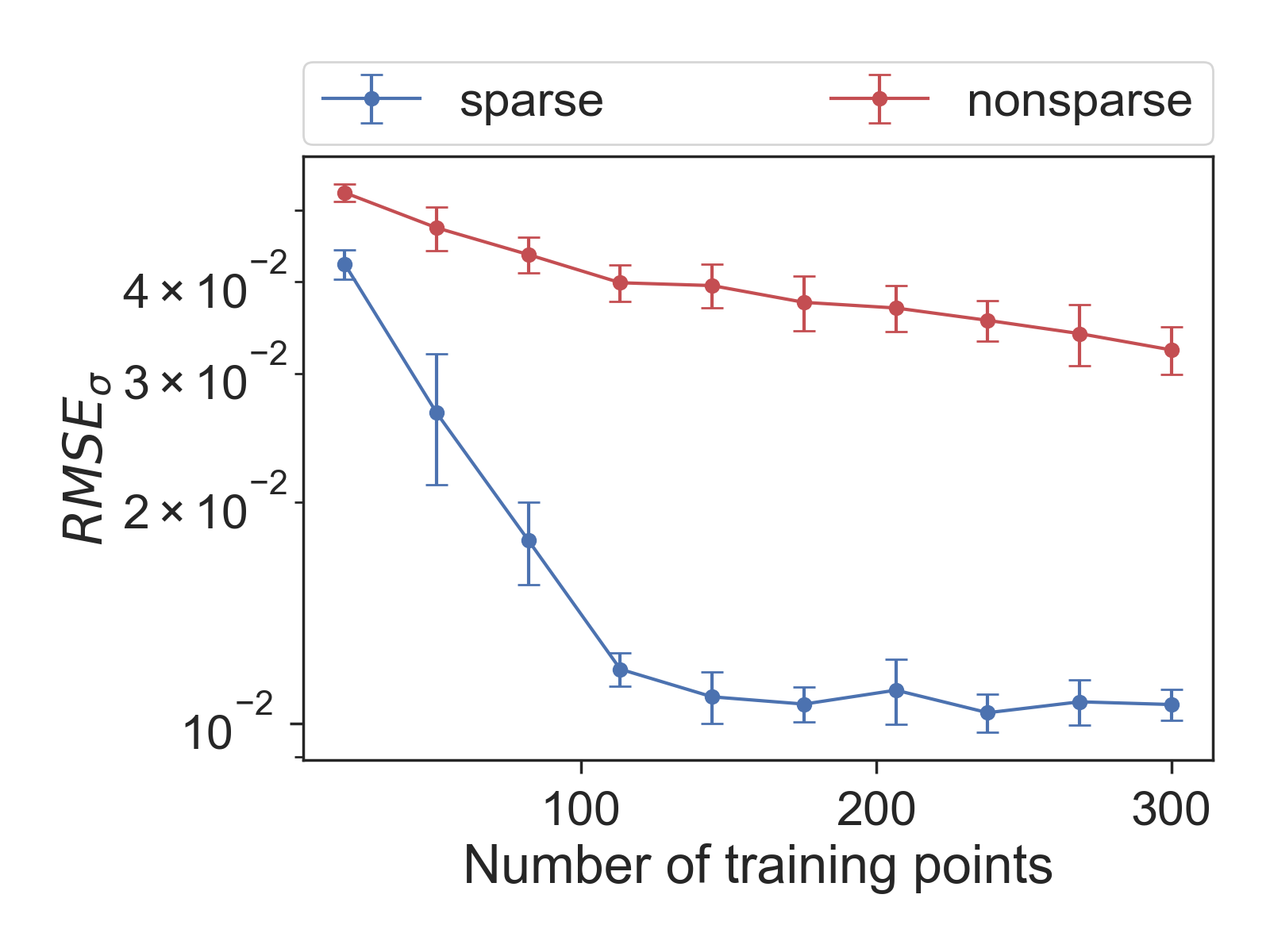}
\caption{RMSE of blade A data fit comparing sparse and non-sparse priors.}
\label{fig:bladeA_sparse_rmse}
\end{figure}

For this study, we construct a BPC model for the blade A dataset. All inputs are normalised to be within the range $[-1, 1]$ and assumed to be independent, with uniform marginals. We seek a quadratic response surface over an isotropic total order polynomial basis of maximum degree 2, resulting in 351 coefficients. The goal is to arrive at the posterior distribution for the coefficients, minimising the predictive error of the posterior mean function, with as few data points as possible, while using the fact that coefficients are expected to be sparse. In \Cref{fig:bladeA_sparse_rmse}, the normalised RMSE on test data is shown for various numbers of training points, comparing the use of the regularised horseshoe prior \eqref{equ:sparse_prior} and a non-sparse prior (a centred Gaussian distribution with covariance $\sigma_\tau \mI$). The hyperparameters are set as $\nu = 25, s = 3, \beta = 0.1$, and $\sigma_\tau$ determined via a hierarchical prior of HalfNormal(1). It is clear that the sparsity-promoting property of the regularised horseshoe prior has improved the predictive accuracy of the fit. 

\subsection{Coregional models}
In the following, we compare the predictive capabilities of a coregional model on multiple functions with separate Bayesian polynomials using uninformative priors. 
\subsubsection{Analytical functions}
\begin{figure}
\centering
\includegraphics[width=0.5\textwidth]{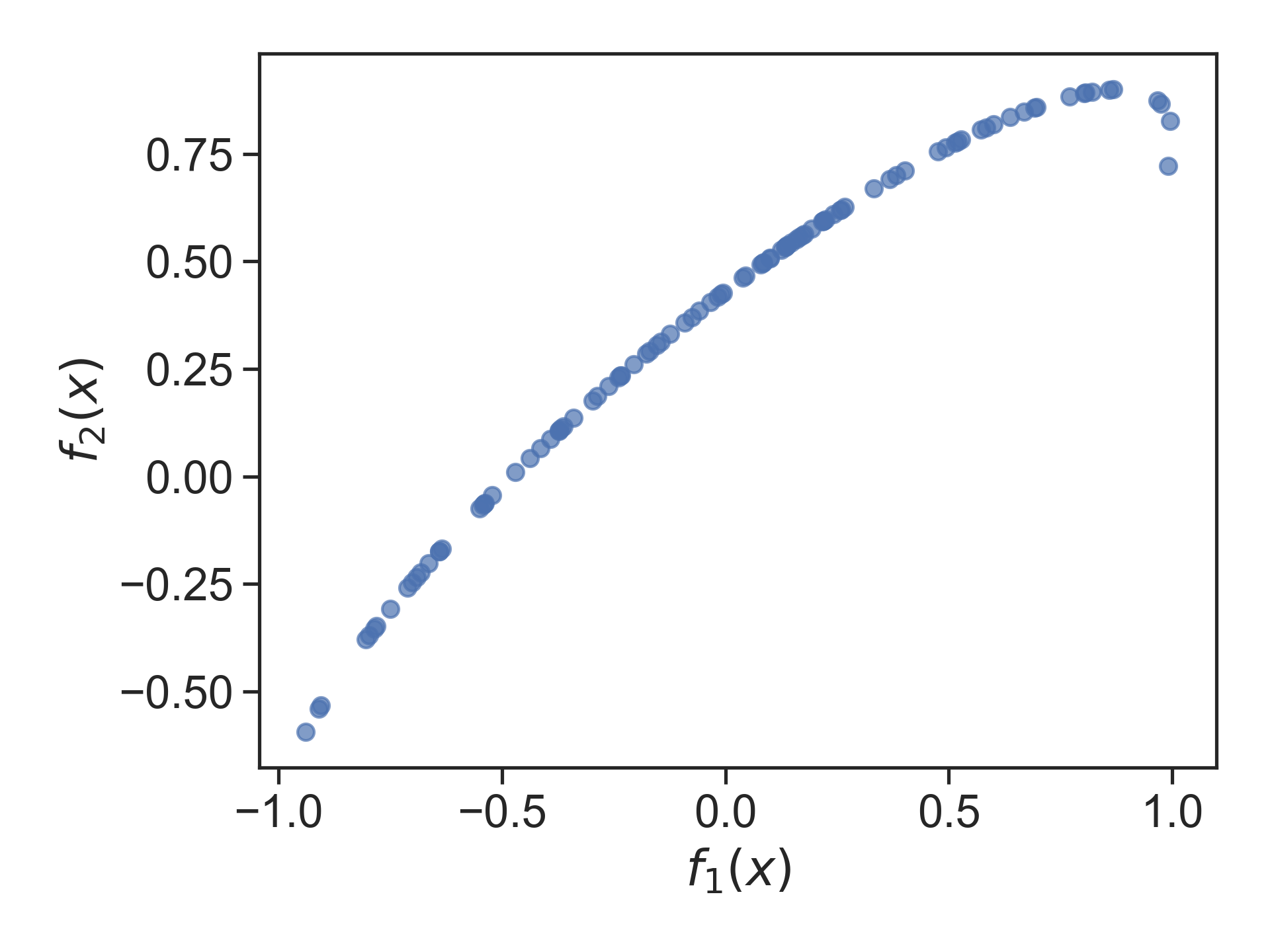}
\caption{Scatterplot showing the functions $f_1, f_2$ evaluated on the same input points. }
\label{fig:simple_func_rel}
\end{figure}

\begin{figure}
\centering
\includegraphics[width=0.8\textwidth]{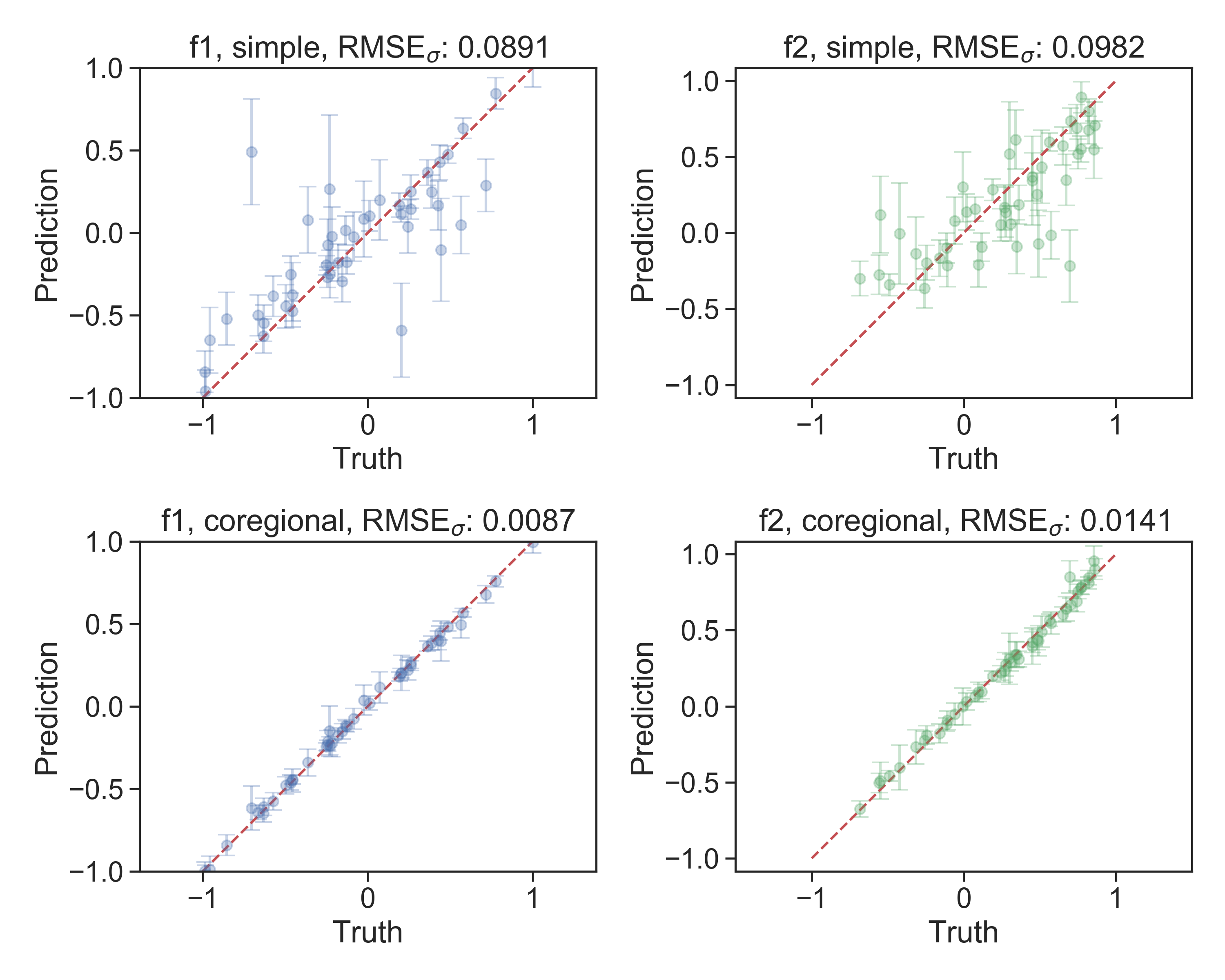}
\caption{Predictions on test points for $f_1, f_2$ plotted against truth. Red dashed lines show where prediction = truth.}
\label{fig:coreg_simple}
\end{figure}

First, consider a pair of analytical functions $f_1, f_2: [-1, 1]^7 \rightarrow \mathbb{R}$, 
\begin{equation}
\begin{split}
f_1(\vx) = \sin \left(\frac{\sum_i x_i}{\sqrt{7}}\right),\\
f_2(\vx) = 0.9\sin \left(\frac{\sum_i x_i}{\sqrt{7}} + 0.5\right).
\end{split}
\end{equation}
Though the functions are distinct, the output points given the same set of inputs are correlated (see \Cref{fig:simple_func_rel}). This characteristic can be exploited with the output correlation structure that is enabled by the coregional model. 
We aim to fit a cubic Legendre polynomial on an isotropic total order basis with maximum degree 3, implying that 120 polynomial coefficients need to be solved for. Using 105 training inputs (different for each output), we fit two models with $\sigma_m = 0.001$:
\begin{itemize}
\item Two Bayesian polynomial models, with prior on the coefficients $p(\boldsymbol{\alpha}) = \mathcal{N}(0, 0.001 \mI)$\footnote{Since we are fitting an underdetermined polynomial, a narrow prior helps regularise the coefficients.}. 
\item A coregional polynomial model, with the priors on hyperparameters $\mA$ and $\mB$ set as described in \eqref{eqn:coreg_hyp}.
\end{itemize}
The posterior mean and standard deviation of the matrix $\mB$ are
\begin{equation}
\begin{bmatrix}
6.592 \pm 1.599 & 6.975 \pm 1.458 \\
6.975 \pm 1.458 & 8.210 \pm 2.031
\end{bmatrix} \times 10^{-4}.
\end{equation}
These values cannot be interpreted directly as output (co)variances owing to the scaling with the values in $\mA$, but the off-diagonal entries clearly show that correlation between the two outputs is incorporated in the predictions. \Cref{fig:coreg_simple} plots the resultant mean predictions of the fitted polynomial for two independent models (simple) and a coregional polynomial model against the true function evaluations at test points, along with the normalised RMSE values on these test points. It can be seen that the coregional model offers a higher predictive accuracy with the same training data when compared with two independent polynomial models.

\subsubsection{Compressor blade efficiency datasets}

Now, we consider the dataset consisting of the non-dimensionalised efficiencies of fan blades A, B and C, as described in \Cref{sub:ex_structural}. Though the CFD simulations of these blades are run at different boundary conditions and have different geometries, the output (normalised) stage efficiencies can be correlated with each other because of the identical design space. In addition, the similarities in the problem setup and physics involved suggest that the efficiencies as functions of the geometries behave similarly for all blades. 

\begin{figure}
\centering
\includegraphics[width=1.0\textwidth]{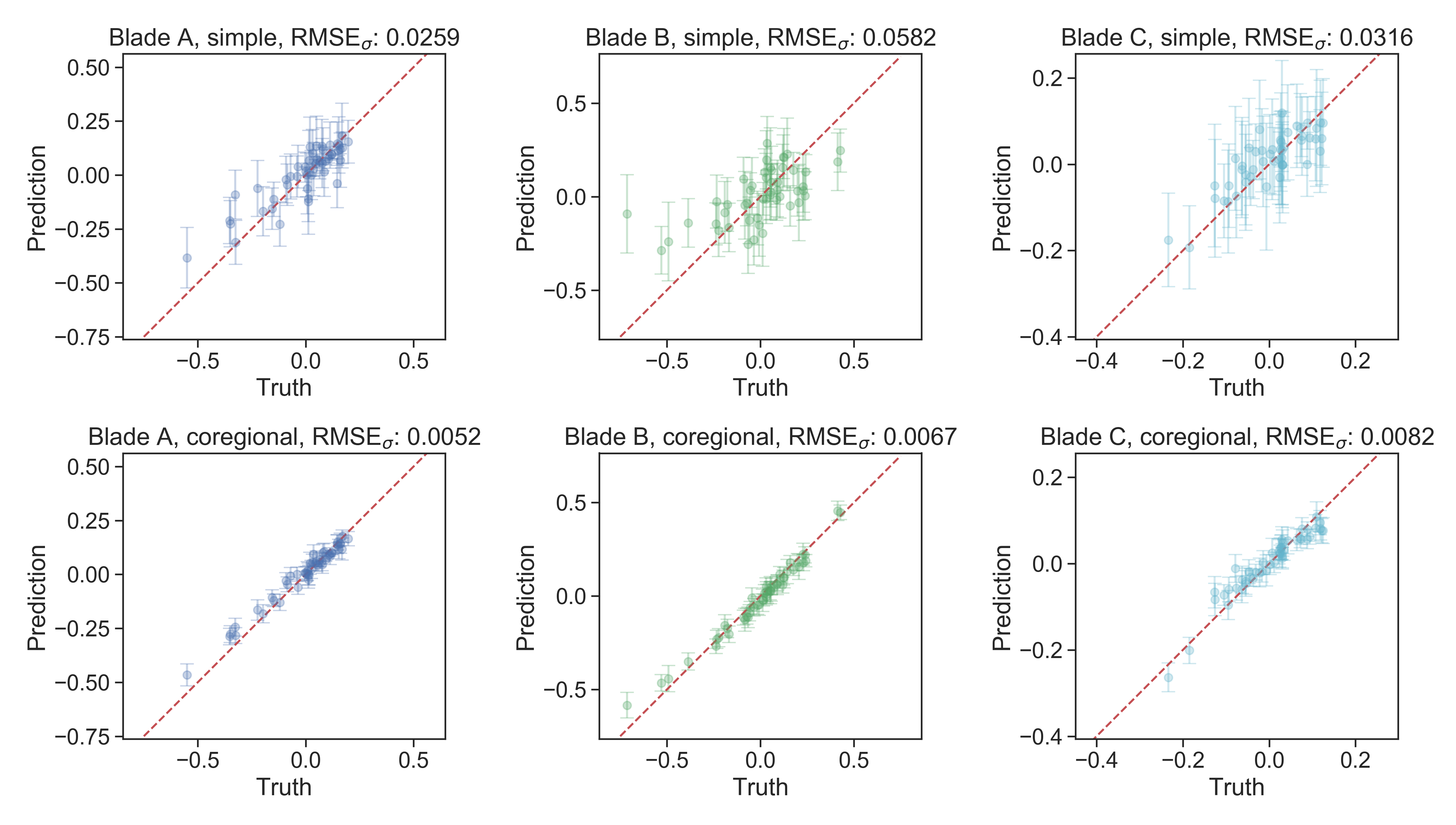}
\caption{Predictions on test points for the compressor blades datasets plotted against truth. Red dashed line shows where prediction = truth.}
\label{fig:coreg_ABC}
\end{figure}

Half of each dataset (274, 155, 273 points respectively) is sampled randomly for training, reserving the other points for testing. A quadratic model on a total order basis with 351 basis terms is fitted for each objective. The same two methods listed above for the analytical functions are compared, where $\sigma_m = 0.001$. \Cref{fig:coreg_ABC} shows the results of both methods on the testing data for all three datasets, showing a clear decrease in the test normalised RMSE, as before, when using the coregional model over independent models. The posterior mean and standard deviation of $\mB$ are
\begin{equation}
\begin{bmatrix}
 24.33 \pm 3.93 &  -26.74 \pm 3.81 & 10.32 \pm 1.60 \\
 -26.74 \pm 3.81 &  33.98 \pm 5.90  & -11.62 \pm 1.93 \\
 10.32 \pm 1.60  & -11.62 \pm 1.93 &  7.629  \pm 1.38
\end{bmatrix} \times 10^{-6},
\end{equation}
which again show significant output correlations between the different efficiency values.

\section{Conclusions}
The focus of this paper is on the reformulation of classical polynomial chaos and approximations from a Bayesian point of view along with its implications. Based on the probabilistic framework of Bayesian polynomial chaos, prior and expert knowledge on the behaviour of the quantities of interest to be approximated can be leveraged in intuitive ways to improve the predictive power of a Bayesian polynomial model. We propose a set of methods that achieves this goal assuming a range of different contexts, including the incorporation of approximate knowledge on linear functionals of the output (e.g., the output mean), the sparsity pattern of the output, and relations between multiple outputs. On a range of numerical examples that include analytical functions and datasets derived from physical simulations and experiments, it is shown that the methods presented in this paper effectively transfer prior knowledge to output predictions, reducing the predictive error on test data. In spite of the variety of existing methods that exploit many facets of polynomial chaos, there yet remains a great expanse of classical theory behind polynomial approximations, such as its rich connections with numerical quadrature and---more recently---dimension reduction (such as \cite{hokanson2018data, wong2020embedded}), we have not covered in this paper. Thus, this paper is intended to initiate discussions on extending the use of orthogonal polynomials and related ideas to a fully probabilistic framework.

\section*{Acknowledgements}
CYW acknowledges financial support from the Cambridge Trust, Jesus College, Cambridge, and the Data-Centric Engineering programme of The Alan Turing Institute. PS was funded through a Rolls-Royce research fellowship. ABD was supported by the Lloyd’s Register Foundation Programme on Data Centric Engineering and by The Alan Turing Institute under the EPSRC grant EP/N510129/1. AS was funded through the Digital Twins in Aeronautics grant as part of the Strategic Priorities Fund EP/T001569/1.

\appendix
\section{Quadratic forms of Gaussians} \label{sec:appen}
Suppose we have an $n$-dimensional Gaussian random vector
\begin{equation}
\boldsymbol{\alpha} \sim \mathcal{N}(\boldsymbol{\mu}, \mSigma),
\end{equation}
where we assume that $\mSigma$ is symmetric positive semi-definite. We are interested in the distribution of a quadratic form defined by
\begin{equation}
\boldsymbol{\alpha}^T \mE \boldsymbol{\alpha}
\end{equation}
for a symmetric $\mE \in \mathbb{R}^{n\times n}$. First consider
\begin{equation}
\boldsymbol{\alpha}^T \mE \boldsymbol{\alpha} = (\boldsymbol{\alpha} - \boldsymbol{\mu})^T \mE  (\boldsymbol{\alpha} - \boldsymbol{\mu}) + 2(\boldsymbol{\alpha} - \boldsymbol{\mu}) ^T \mE \boldsymbol{\mu} + \boldsymbol{\mu}^T \mE \boldsymbol{\mu}.
\end{equation}
The first term can be written as
\begin{equation}
\begin{split}
(\boldsymbol{\alpha} - \boldsymbol{\mu})^T \mE  (\boldsymbol{\alpha} - \boldsymbol{\mu}) &= (\sqrt{\mSigma}\va)^T \mE (\sqrt{\mSigma}\va)\\
&= \va^T \sqrt{\mSigma}^T \mE \sqrt{\mSigma} \va,
\end{split}
\end{equation}
where $\va \sim \mathcal{N}(0, \mI)$. That is, this term can be expressed as a weighted sum of chi-squared distributed variables. The second term is a linear combination of centred Gaussian variables $(\boldsymbol{\alpha} - \boldsymbol{\mu})$, and the last term is a constant. Overall, this constitutes a linear combination of non-centred chi-squared variables.

\bibliographystyle{elsarticle-num}
\bibliography{references}

\end{document}